\documentclass[a4paper,11pt]{article}
\usepackage{pos}

\title{Improving the CMS High Level Trigger tracking at the HL-LHC with novel and evolved heterogeneous algorithms}
\ShortTitle{Improving the CMS High Level Trigger tracking at the HL-LHC}

\author*[a]{Mario Masciovecchio}
\onbehalf{on behalf of the CMS Collaboration}

\affiliation[a]{University of California, San Diego,\\
9500 Gilman Dr., La Jolla, CA 92093, United States of America}

\emailAdd{mario.masciovecchio@cern.ch}

\abstract{Charged particle track reconstruction is one of the heaviest computational tasks in the event reconstruction chain at Large Hadron Collider (LHC) experiments. Furthermore, projections for the High Luminosity LHC (HL-LHC) show that the required computing resources for single-threaded CPU algorithms will exceed those that are expected to be available. It follows that experiments at the HL-LHC will need to employ novel and evolved track reconstruction algorithms, within heterogeneous computing systems that include many-core CPUs as well as GPUs, in the attempt to maximize the computational performance while retaining the best possible reconstruction efficiency. In the context of the Compact Muon Solenoid (CMS) High Level Trigger (HLT) at the HL-LHC, the mkFit algorithm, already in use for the CMS track reconstruction during the LHC Run 3, will exploit its parallelized and vectorized nature on CPUs to perform pattern recognition using seed tracks produced with algorithms that are designed to be fully parallelizable and hardware agnostic, thus suitable for heterogeneous systems: the Patatrack and the Line Segment Tracking (LST) algorithms. Patatrack is an established algorithm, already used for the CMS pixel track reconstruction at HLT during the Run 3 of the LHC, while LST is a novel algorithm, recently integrated in the CMS software, targeting the reconstruction of tracks in the outer tracker of the HL-LHC CMS detector. The state-of-the-art performance for the CMS HLT track reconstruction at the HL-LHC is presented, obtained using the combination of the mkFit, Patatrack, and LST algorithms, that in turn use machine-learning (ML) techniques such as deep neural networks and multi-objective particle swarm optimization to suppress duplicate and misreconstructed tracks. Prospects of further improvements are also presented, with a focus on the usage of ML techniques for track reconstruction at CMS.}

\FullConference{23rd International Workshop on Advanced Computing and Analysis Techniques in Physics Research (ACAT2025)\\
8–12 September 2025\\
Hamburg, Germany\\}

\begin{document}
\maketitle


\section{Introduction}

The High-Luminosity LHC (HL-LHC) will represent the start of a new era
for particle physics, delivering an integrated luminosity larger by one order of magnitude with respect to previous LHC running periods.
The increased dataset will extend the discovery potential for physics beyond the standard model (SM) and significantly improve the precision of SM measurements.
However, these opportunities come with substantial computational challenges.
The HL-LHC is expected to operate with up to 200 simultaneous
proton-proton interactions per bunch crossing (pileup, PU), resulting in detector
occupancies far exceeding those encountered during the Run~3 of the LHC.
The larger number of detector hits leads to increased combinatorics for track
finding and consequently a significant growth in reconstruction
complexity.

Charged-particle tracking is one of the most computationally intensive
tasks of the CMS event reconstruction sequence,
both offline and at the High Level Trigger (HLT).
In particular at HLT, one must reconstruct tracks while
satisfying strict latency constraints, retaining the best possible
physics performance at the same time.
Without substantial improvements to reconstruction
algorithms and computing models the resources required to process
HL-LHC data would increase beyond sustainable levels.
To address these challenges, CMS is pursuing several, complementary
developments aimed at improving both performance and scalability of
future tracking, including tracking at HLT.
These efforts include heterogeneous CPU/GPU reconstruction 
with Patatrack~\cite{CMS-DP-2022-014,CMS-DP-2024-083},
parallelized Kalman-filter tracking with mkFit~\cite{CMS-DP-2022-018},
novel tracking algorithms optimized for the phase-2 tracker geometry
such as the Line Segment Tracking (LST) algorithm~\cite{CMS-DP-2023-019,CMS-DP-2024-014},
and machine-learning-enhanced candidate selection~\cite{CMS-DP-2023-075}.

In the following, recent developments are presented in the context of the future CMS HLT tracking strategy for HL-LHC operations.


\section{Track reconstruction at the CMS HLT at HL-LHC}
\label{previous}

\begin{figure}[htbp]
\centering
\includegraphics[width=0.95\textwidth]{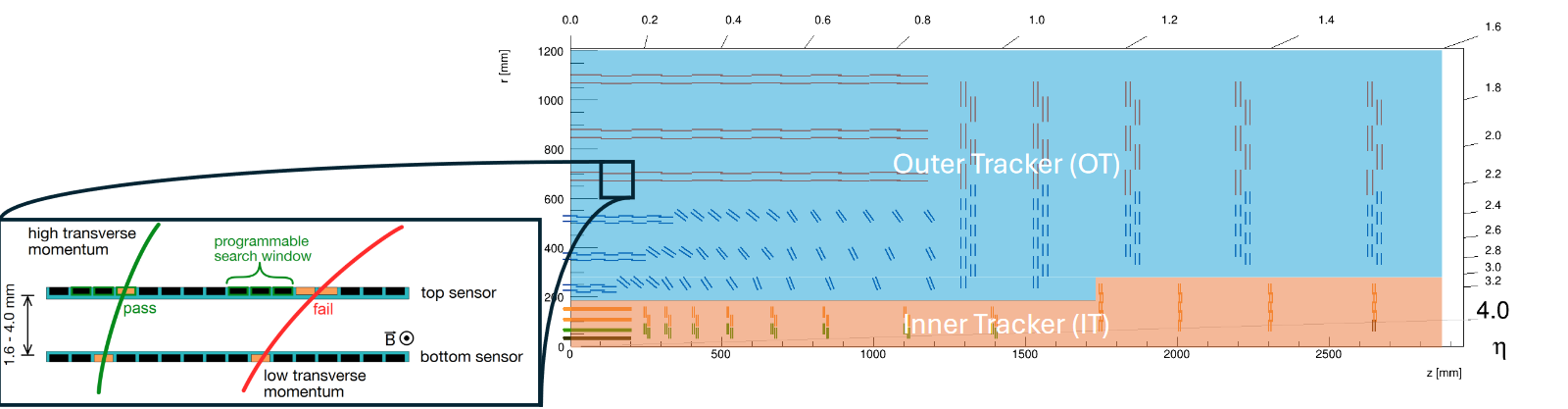}
\caption{
Layout of the CMS phase-2 tracker detector, showing the Inner Tracker
and Outer Tracker subsystems, with a detail showing the formation of mini-doublets (MD).
}
\label{fig:tracker}
\end{figure}

The CMS phase-2 tracker, depicted in Fig.~\ref{fig:tracker}
has been designed to ensure efficient track
reconstruction under the extreme HL-LHC conditions.
The detector consists of an Inner Tracker (IT),
composed entirely of silicon pixel sensors, and an Outer Tracker (OT), composed of stacked strip and macro-pixel modules.
A distinctive feature of the OT is the possibility 
of forming ``mini-doublets'',
pairs of hits measured on adjacent sensors within the same detector module. These objects provide an approximate transverse momentum ($p_\mathrm{T}$) estimate and can be exploited by tracking algorithms.

The baseline HL-LHC HLT charged-particle track reconstruction sequence consists,
at the time of this presentation, of two iterations, seeded using four- and three-hit pixel tracks, respectively.
Both iterations employ the traditional ``Combinatorial Kalman Filter''
(CKF) for pattern recognition and track building.
The CKF algorithm has demonstrated excellent reconstruction performance throughout CMS operations and remains the reference tracking algorithm for many applications.
However, the inherently sequential nature of Kalman-filter track
building limits opportunities for parallelization and efficient
usage of accelerators.
This motivates the investigation of alternative tracking approaches
that are better suited to modern heterogeneous computing architectures and
the increased occupancy conditions expected at the HL-LHC.

The performance metrics used throughout this work are the
tracking efficiency, fake rate, duplicate rate, and computational
performance.
A reconstructed track is considered associated with a
simulated particle when more than 75\% of its hits originate from that
particle.
The tracking efficiency is defined as the fraction of simulated particles
that are associated to at least one reconstructed track.
Fake rate is defined as the fraction of reconstructed tracks that are not associated to any simulated particle. Duplicate rate is defined as the fraction of reconstructed tracks associated multiple times to the same simulated particles.
Performance is measured using a $t\bar{t}$ simulated sample with 200 PU at a center-of-mass energy of 14~TeV, with the upgraded phase-2 detector geometry.


\section{Novel approaches for the HL-LHC HLT track reconstruction}

To address the challenges posed by HL-LHC PU conditions,
CMS is developing several, complementary tracking approaches that exploit
modern computing architectures and novel reconstruction strategies.
These algorithms target different stages of the track reconstruction chain
and are designed to improve both computational performance and physics
capabilities.


\subsection{Patatrack: heterogeneous pixel tracking}

Patatrack is a heterogeneous reconstruction framework developed within
CMS to exploit hardware accelerators, primarily Graphics Processing
Units (GPUs),
for pixel track reconstruction~\cite{CMS-DP-2022-014,CMS-DP-2024-083}.
The algorithm performs pixel clustering, hit reconstruction, track
seeding, building and fitting, and vertex reconstruction using massively
parallel workflows.
Since most operations involve numerous independent
detector hits and track candidates, they are well suited
to execution on GPUs.

Patatrack was already successfully deployed
in the CMS Run-3 HLT~\cite{CMS-DP-2022-014}, relying on a multi-objective particle swarm optimization of the reconstruction parameters in order to maximize the efficiency while suppressing duplicate and misreconstructed tracks~\cite{CMS-DP-2024-084}.
In that context, the reconstructed pixel tracks are
subsequently used to seed later tracking tasks running on conventional CPUs.
For HL-LHC HLT applications, Patatrack provides a natural solution for
handling the increased event complexity.
Current (at the time of this presentation) developments
focus on extending the framework to future
detector geometries and further improving integration within
heterogeneous trigger farms~\cite{CMS-DP-2024-083}.


\subsection{mkFit: parallelized Kalman-filter tracking}

Although accelerators provide substantial computational power,
efficient utilization of modern multi-core CPUs remains equally
important. The mkFit project addresses this challenge through a
 redesign of Kalman-filter track reconstruction~\cite{CMS-DP-2022-018}.
The algorithm preserves the well-established Kalman-filter formalism
used throughout CMS tracking while introducing extensive thread-level
parallelism and SIMD vectorization. This is achieved through the
\texttt{Matriplex} data layout, which enables simultaneous processing of
multiple track candidates.
Unlike traditional CKF reconstruction,
which contains many inherently sequential components, mkFit is
explicitly optimized for contemporary processor architectures.
The resulting implementation improves processor utilization and event
throughput
while retaining excellent physics performance~\cite{CMS-DP-2022-018}.

Since Run~3, mkFit has been integrated into CMS tracking
reconstruction and is responsible for the majority of reconstructed
offline tracks. Since 2025, it was also deployed in production at HLT.
Ongoing (at the time of this presentation) HL-LHC studies investigate its deployment within future HLT tracking workflows,
where it offers a scalable CPU-based solution for track reconstruction~\cite{CMS-DP-2025-051}.


\subsection{Line Segment Tracking}

Line Segment Tracking (LST) is a highly parallel tracking algorithm
specifically designed to exploit the geometry of the CMS Phase-2
Outer Tracker \cite{CMS-DP-2023-019,CMS-DP-2024-014}.
The algorithm follows a fundamentally different strategy from
Kalman-filter-based approaches. Rather than extending trajectories
layer by layer, LST constructs tracks hierarchically from localized
detector objects.

\begin{figure}[htbp]
\centering
\includegraphics[width=0.32\textwidth]{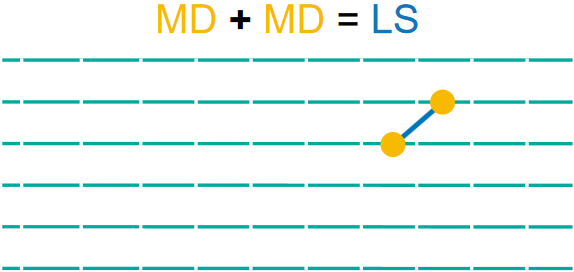}
\includegraphics[width=0.32\textwidth]{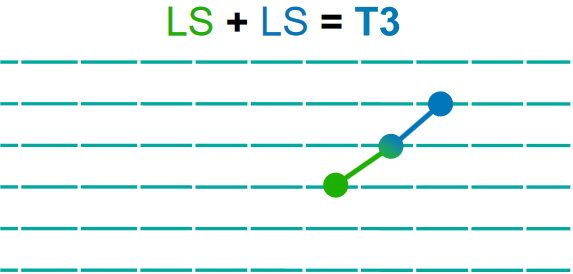}
\includegraphics[width=0.32\textwidth]{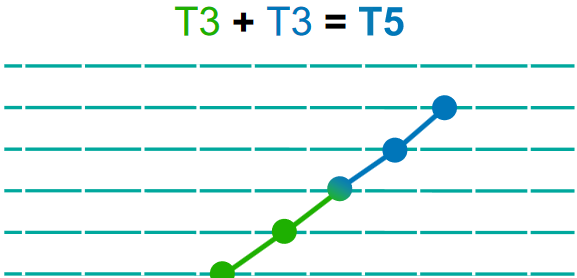}
\caption{
Hierarchical object construction in the Line Segment Tracking
algorithm. (Left) Mini-doublets (MD) are combined into Line Segments (LS); (center) LSs are combined into triplets (T3);
and (right) T3s are combined into quintuplets (T5). Then, pixel-track information is associated to T5s and T3s, to form pT5
and pT3 candidates, respectively.
}
\label{fig:lstHierarchy}
\end{figure}

The reconstruction begins with the formation of mini-doublets (MD),
pairs of correlated hits originating from a single Outer Tracker
module. Mini-doublets are linked into larger objects called Line
Segments (LS), which are subsequently combined into triplets (T3) and
quintuplets (T5), as shown in Fig.~\ref{fig:lstHierarchy}.
Information from Inner Tracker pixel tracks is then 
associated to T5s and T3s, to form pT3 and pT5 candidates, respectively.

Because all reconstruction stages involve linking relatively local
objects, the algorithm exposes a very large degree of parallelism and
is naturally suitable for execution on CPUs or GPUs. Additional
geometrical selections and machine-learning-based classifiers are used
to suppress fake combinations and control combinatorial growth under
high-PU conditions
\cite{CMS-DP-2023-019,CMS-DP-2023-075,CMS-DP-2024-014}.

A particularly attractive feature of LST is its ability to reconstruct
tracks with large transverse impact parameters. As a consequence, the
algorithm significantly enhances sensitivity to long-lived particle
signatures and other exotic topologies that can be challenging for
traditional prompt-tracking algorithms
\cite{CMS-DP-2023-019,CMS-DP-2024-014}.

At the time of this presentation, the LST algorithm is being developed
with the goal to improve the track reconstruction at HLT at the HL-LHC~\cite{CMS-DP-2025-051}.


\section{Towards a new HL-LHC HLT tracking baseline}
\label{sec:new-baseline}

A novel baseline for HLT tracking at HL-LHC shall leverage
all available developments.
In this presentation, the previous baseline described
in Sec.~\ref{previous} is compared 
to alternate options that leverage mkFit and LST recent developments~\cite{CMS-DP-2025-051}.
In one case (``LST+CKF''), LST is used in the first iteration.
In the second one (``LST+mkFit''), mkFit is also used, replacing CKF in the second iteration.
%

\begin{figure}[htbp]
\centering
\includegraphics[width=0.42\textwidth]{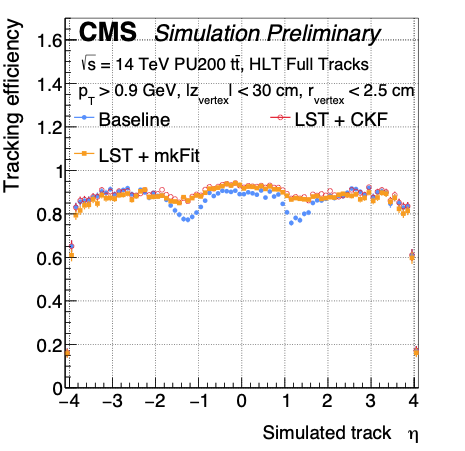}
\includegraphics[width=0.42\textwidth]{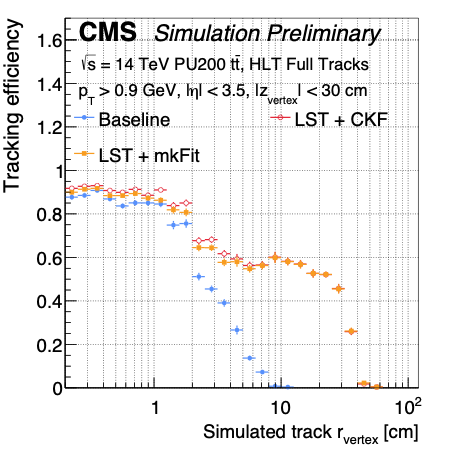}
\caption{
Track reconstruction efficiency for previous baseline (blue), LST+CKF (red) and LST+mkFit (orange) configurations, as a function of the simulated track pseudorapidity (left)
and transverse displacement from the hard-interaction vertex (right).
From Ref.~\cite{CMS-DP-2025-051}.
}
\label{fig:eff}
\end{figure}

\begin{figure}[htbp]
\centering
\includegraphics[width=0.42\textwidth]{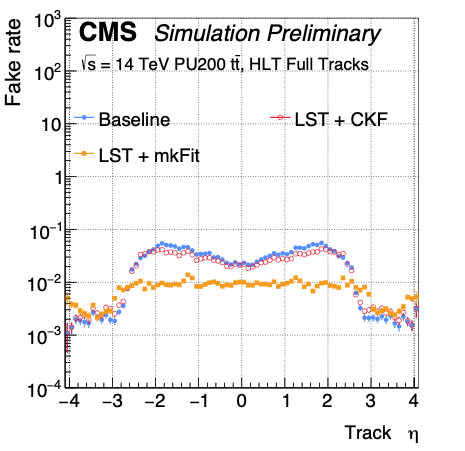}
\includegraphics[width=0.42\textwidth]{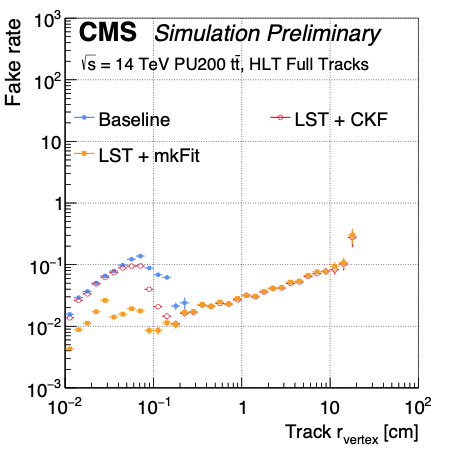}
\caption{
Track reconstruction fake rate for previous baseline (blue), LST+CKF (red) and LST+mkFit (orange) configurations,
as a function of the track pseudorapidity (left) 
and transverse displacement from the hard-interaction vertex (right).
From Ref.~\cite{CMS-DP-2025-051}.
}
\label{fig:fake}
\end{figure}

The tracking efficiency (fake rate) for all three configurations is shown in Fig.~\ref{fig:eff} (Fig.~\ref{fig:fake}).
On the one hand, using LST enables a significant improvement of the tracking efficiency, especially at large displacement from the hard-interaction vertex, up to 40~cm.
On the other hand, when mkFit is used a reduction in the number of fake tracks by about three times is observed. Concurrently, the (CPU) track building time is significantly reduced when mkFit is used.

\section{Using ML to enhance tracking: the LST example}

\begin{figure}[htbp]
\centering
\includegraphics[width=0.32\textwidth]{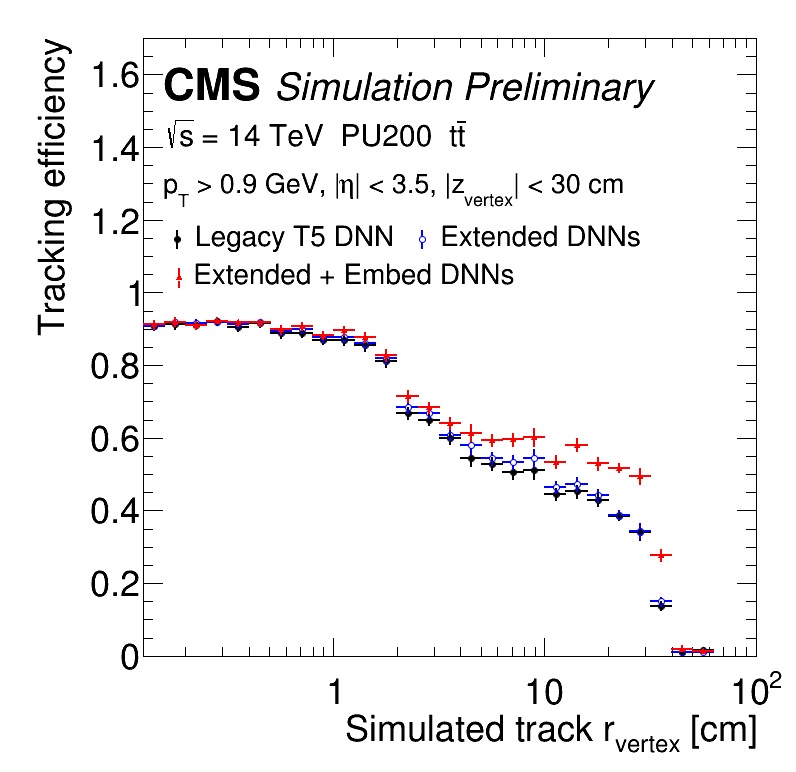}
\includegraphics[width=0.32\textwidth]{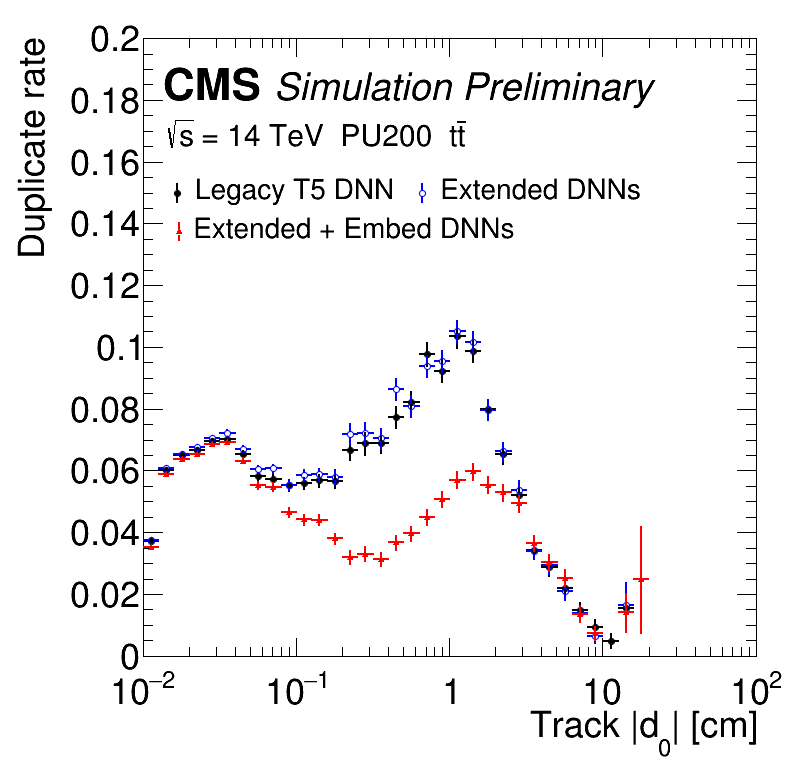}
\includegraphics[width=0.32\textwidth]{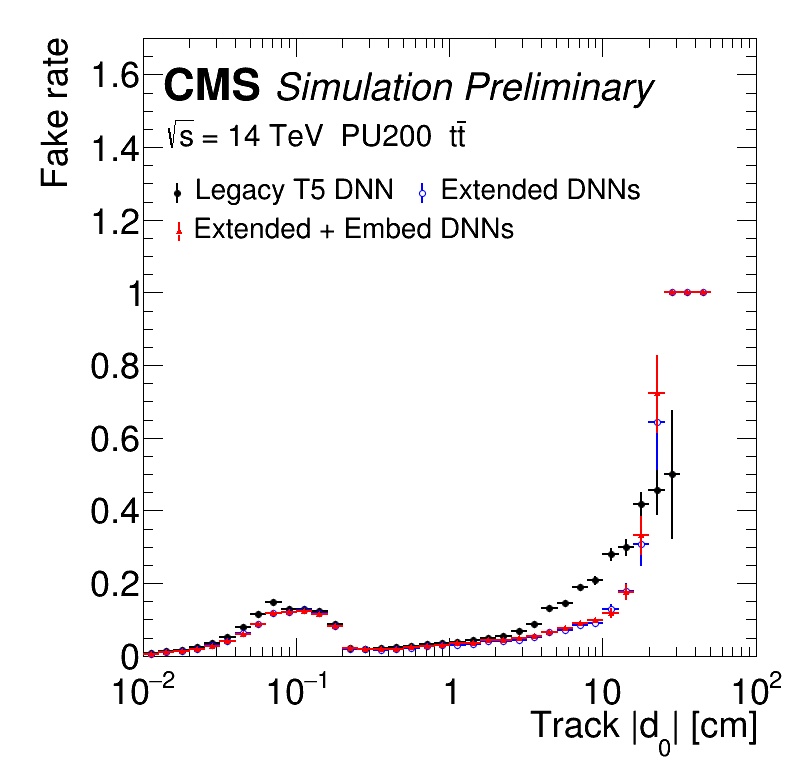}
\caption{
Track reconstruction efficiency (left), duplicate rate (center)
and fake rate (right) for LST candidates as a function of the track transverse displacement from the hard-interaction vertex,
for the ``legacy'' (i.e., previous) configuration (black), using extended DNN selections (blue), and using both extended DNN selections and track embeddings for duplicate removal (red).
From Ref.~\cite{CMS-DP-2025-048}.
}
\label{fig:ml-lst}
\end{figure}

The performance of LST, and in turn of the overall HLT tracking at the HL-LHC, was recently improved
through additional ML-based candidate selections~\cite{CMS-DP-2025-048}.
These improvements are included in the results presented in Sec.~\ref{sec:new-baseline} and in Ref.~\cite{CMS-DP-2025-051}.
In particular, new ML classifiers were applied to intermediate LST objects in order
to reject combinatorial candidates that survived the geometrical
selection criteria.
While the efficiency of genuine track reconstruction is preserved, 
the candidate purity is improved 
and the number of objects propagated through subsequent reconstruction stages is reduced.
The achieved physics performance is shown in Fig.~\ref{fig:ml-lst}.
By reducing the number of low-quality candidates,
the enhanced configuration also achieves a significant reduction in event (GPU) processing time.

\section{Summary and outlook}

The HL-LHC will present significant challenges for CMS HLT tracking,
requiring new approaches capable of maintaining physics performance
within realistic computing budgets. This work presented a revised HL-LHC HLT tracking strategy combining
mkFit and LST. These developments provide improvements in both
reconstruction performance and computational scalability.

Future studies will focus on further extending the use of ML techniques,
accelerating track seeding through heterogeneous approaches such as Patatrack,
and leveraging mkFit for additional tracking tasks, including track fitting.
The improved performance of these algorithms also motivates investigations
of faster HLT workflows, potentially reducing the number of
tracking iterations required during event reconstruction.

These developments represent crucial steps toward a new generation
of CMS HLT tracking reconstruction for phase-2 operations.


\acknowledgments{This work was supported by the National Science Foundation under Cooperative Agreements OAC-1836650 and PHY-2323298.}



\end{document}